# Predictions for the 4 GeV TJNAF inclusive electron scattering experiment and for FSI effects in EMC ratios

A.S. Rinat and M.F. Taragin

*Department of Particle Physics, Weizmann Institute of Science, Rehovot 76100, Israel*

(February 6, 1997)

## Abstract

We express nuclear structure functions $F_i^A$ as generalized convolutions of the structure function of a nucleon and of a nucleus, composed of point-nucleons. In computations of the latter we include Final State Interactions and results for $F_2^A$ are compared with a few directly measured data on C and Fe. The above $F_i^A$ are primarily used for predictions of the TJNAF 89-008 inclusive scattering experiment of 4 GeV electrons on various targets. Those cover a broad angular, and correspondingly wide $x, Q^2$ range, where the nucleon-inelastic part dominates large sections of the covered kinematics. The same model has been applied to the study of hitherto neglected Final State Interaction effects in the nuclear component in EMC ratios in the region $0.85 \gtrsim x \gtrsim 0.25$.

# I. INTRODUCTION.

In the following we discuss two topics on inclusive scattering of electrons from nuclear targets. The first is the TJNAF 89-009 experiment at $E$=4 GeV beam energy on a variety of targets, mainly Fe [1]. The angular range $15° \lesssim \theta \lesssim 135°$ covers $1 \lesssim Q(\text{GeV}^2) \lesssim 7.1$; $0.55 \lesssim x \lesssim 4.15$ and extends the older NA3 data for $E$=2-4 GeV [2], as well as for the more recent NA18 data [3] for relatively wide $Q^2$ but limited $x \approx 1$. The TJNAF results which are currently been analysed, will show progressive dominance of the inelastic part of cross sections for scattering angles as low as $\theta \gtrsim 60°$, even for energy losses $\nu$ on the elastic side of the quasi-elastic peak. Using the same model we address as a second topic Final State Interaction effects (FSI) in the purely nuclear component in the EMC ratio.

In Section 2 we review our model for the structure functions of nuclei $F_2^A(x, Q^2)$ which is argued to be valid for large $Q^2$ and treat separately components coming from elastic (NE) and inelastic (NI) parts of the nucleon structure function $F_i^N$. Those are predicted to be barely dependent on all but the lightest mass numbers, which, when substantiated, would considerably limit the information one may extract from data on different targets with $A \gtrsim 12$. In Section 3 we compare computed $F_2^A$ with scarce, directly measured data [3,4]. In addition we present predictions for inclusive cross sections corresponding to the kinematics of the TJNAF 89-008 experiment. In Section 4 we report on FSI effects in the nuclear component of the EMC ratio, which appear to improve the agreement with data in the region $x \lesssim x_{min}$. We draw attention to non-negligible NE contributions, growing with increasing $x \gtrsim 0.75$ and decreasing with $Q^2$.

# II. A CONVOLUTION MODEL FOR NUCLEAR STRUCTURE FUNCTIONS.

We consider the cross section per nucleon for inclusive scattering of electrons from nuclear targets $A(Z, N)$

$$\frac{d^2\sigma_{eA}}{d\Omega d\nu} = \frac{1}{A}\sigma_{\text{M}}\left[W_2^A(\nu, Q^2) + 2\text{tg}^2(\theta/2)W_1^A(\nu, Q^2)\right] \qquad (1a)$$

$$= \frac{1}{A}\frac{4\alpha^2}{Q^4}xE(u^{-1}-1)\left[\left(1-u-\frac{Mxu}{2E}\right)\frac{F_2^A(x,Q^2)}{x}+u^2F_1^A(x,Q^2)\right] \tag{1b}$$

$q,\nu$ are the components of the 4-momentum transfer and $Q^2 = q^2 - \nu^2$. $u = \nu/E$ and $\sigma_M = \frac{4\alpha^2}{Q^4}[E(1-u)]^2\cos^2(\theta/2)$, the Mott cross section. In a standard notation $F_1^A = MW_1^A, F_2^A = \nu W_2^A$ are nuclear structure functions per nucleon. Those depend on a pair of kinematic variables, for instance $(q,\nu), (\nu,Q^2)$, or $(x,Q^2)$. We use a Bjorken variable $x = Q^2/2M\nu$ for the nucleus in terms of the nucleon mass $M$ with range $0 \le x \le M_A/M \lesssim A$.

For the nuclear structure functions in (1) we have previously suggested a generalized convolution [5]

$$W_i^A(q,\nu) = \int d\nu' w^{N/A}(q, \nu - \nu' + \mathcal{E}_{\boldsymbol{q}})\left[\frac{Z}{A}W_i^p(q,\nu') + \frac{N}{A}W_i^n(q,\nu')\right] \tag{2}$$

with $\mathcal{E}_{\boldsymbol{q}}$ a nucleon recoil energy. $w^{N/A}$ is the structure function of a fictitious nucleus composed of point particles. It shall be called the *point − nucleon* (PN) nuclear structure function, to be distinguished from the *total* nuclear structure function $W_i^A$ ($F_i^A$). $W_i^N$ ($F_i^N$) refer to isolated nucleons and emphatically not to nucleons in a medium with off-shell kinematics.

The simplest case for which a convolution such as (2) holds, is a composite system with at least two modes having distinctly different excitation energies, described by the Born-Oppenheimer approximation. An example is a liquid of $H_2$ molecules, which was studied more than 30 years ago. The response (structure function) of the system is a convolution of the same for the translational, rotational, vibrational and electronic modes [6].

A second example relevant to our topic is a non-relativistic (NR) quark cluster model with, optionally, a different treatment of interactions between quarks inside the same, and between quarks in different bags. The interaction between quarks inside one nucleon need not be specified, because all information required in (2) is the nucleon structure functions, which are taken from experiment. In contrast, one requires in principle interactions between quarks in different bags. In a particular version we replaced this total interaction between quarks in two bags by an effective one, acting on their centers of mass, i.e. by a conventional $NN$ interaction [5]. At this point one obtains a description, quite similar to that of liquid H$_2$.

A relativistic extension of (2) can only be postulated and we refer to [5] for a series of, partly heuristic steps which lead to

$$F_i^A(x,Q^2) = \int_x^A \frac{dz}{z^{2-i}} f^{N/A}(z,Q^2) \left[ \frac{Z}{A} F_i^p\left(\frac{x}{z},Q^2\right) + \frac{N}{A} F_i^n\left(\frac{x}{z},Q^2\right) \right] \tag{3}$$

The above describes quark distributions in nuclei, where the same for nucleons is spread by PN nuclear structure functions.

The variables $x, z$ in (3) are, strictly speaking, momentum fractions of quarks, respectively nucleons in nuclei [7], which only in the $Q^2 \to \infty$ limit coincide with the kinematic Bjorken variables. However, we shall assume that it is permissible to make a similar identification in (5) for large, finite $Q^2$.

Versions or approximations of (3) have been discussed before. The best known case is a relativistic Plane Wave Impulse Approximation for the total structure function, defined to be (3) with the PN structure function $f^{N/A}$ in the PWIA, in turn related to the nucleon spectral function. An encumbering feature there is the fact that $F_i^N$ are by construction form factors appropriate to a knocked-out nucleon off its mass-shell [8–11].

Another example which leads to (3) is the meson-quark coupling model where, quite similar to the above quark-cluster model, one distinguishes between quarks inside a bag and unrelated couplings of those quarks to $q\bar{q}$ scalar and vector mesons [12]. The latter are of course related to one-boson exchange components of the effective interaction $V_{NN}$.

We return to the nuclear structure functions $F_i^A$ in (3) and present here a mere outline of a calculation, starting from an essentially NR calculation of PN nuclear structure functions $w_{N/A}$ in Eq. (2) [13]. Consider first the reduced response in the formulation by Gersch, Rodriguez and Smith (GRS) [14]

$$\phi^{N/A}(q,y) = (q/M) w^{N/A}(q,\nu) \tag{4a}$$

$$\lim_{q \to \infty} \phi^{N/A}(q,y) = (4\pi^2)^{-1} \int_{|y|}^{\infty} dp\, p\, n(p) \tag{4b}$$

$$y \to y^{nr} = (M/q)(\nu - q^2/2m), \tag{4c}$$

with $y^{nr}$ the non-relativistic GRS-West scaling variable [14]. Eq. (4b) is the asymptotic limit of $\phi$, expressed in terms of the single nucleon momentum distribution $n(p)$. Additional Final State Interaction (FSI) corrections are conveniently introduced in the Fourier transform of the PN structure function

$$\tilde{\phi}^{N/A}(q,s) = \int ds e^{isy} \phi^{N/A}(q,y) = \int ds e^{isy} \int \frac{d\boldsymbol{p}}{2\pi^3} n(p) R(q, y - \boldsymbol{p}\hat{\boldsymbol{q}}) \tag{5a}$$

$$\tilde{R}(q,s) = \int d\boldsymbol{r} \rho(r) \exp[\tilde{\Omega}(q,\boldsymbol{r},s)], \tag{5b}$$

with $\rho(r)$ the single nucleon density. The functions $\tilde{R}$ or the FSI phase $\tilde{\Omega}$ formally contain all FSI effects and can only be calculated in some approximation. As before we retain binary collisions (BCA) which account for repeated collisions between the struck and any arbitrary core nucleon. The BCA contributions are in (5b) written in a cumulant representation with a FSI phase [13]

$$\tilde{\Omega}(q, \boldsymbol{r}_1, s) \to \tilde{\Omega}_2(q, \boldsymbol{r}_1, s) = \int \rho(\boldsymbol{r}_1 - \boldsymbol{r}) \zeta_2(\boldsymbol{r}, s) \tilde{\Gamma}(q, \boldsymbol{r}, s) \tag{6}$$

For $\zeta_2$ we chose $\zeta_2 \approx \sqrt{g(\boldsymbol{r})g(\boldsymbol{r} - s\hat{\boldsymbol{q}})}$, with $g$ the pair distribution function [14]. The above collisions generally occur off-shell, causing also the characteristic profile $\tilde{\Gamma}$ to be off-shell. For short-range interactions the latter and its on-shell analog are approximately related by [13,15]

$$\tilde{\Gamma}(q, \boldsymbol{r}, s) \approx \theta(s - z) \theta(z) \Gamma(q, b)$$
$$\Gamma(q, b) \approx \frac{1}{2} \sigma_q^{tot} (1 - i\tau_q) A_q(b)$$
$$A_q(b) \approx \frac{Q_0^2(q)}{4\pi} \exp{-\frac{1}{2} b^2 Q_0^2(q)} \tag{7}$$

The latter profile may be conveniently parameterized in terms of $NN$ scattering parameters, as are the total cross section $\sigma_q$, the ratio $\tau_q$ of the real to imaginary part of the forward elastic scattering amplitude and some range parameter $Q_0(q)$ in the latter. Using (7) in (6) one finds for the FSI phase function in the BCA [13]

$$\tilde{\Omega}(q, \boldsymbol{r}_1, s) = \int \rho(\boldsymbol{r}_1 - \boldsymbol{r}) \zeta_2(\boldsymbol{r}, s) \left\langle \tilde{\Gamma}(q, \boldsymbol{r}, s) \right\rangle$$

$$\approx -\left\langle \frac{1}{2}\sigma_q^{tot}(1-i\tau_q)\int d^2\mathbf{b}A_q(b)\right\rangle$$
$$\times \int_0^s dz\rho(\mathbf{r}_1-\mathbf{r})\zeta_2(b,z,s)[1-s\delta(s-z)], \tag{8}$$

where brackets indicate weighted averages of $pp$ and $pn$ parameters.

Eqs. (4)-(8) contain the essential features of a NR PN structure function in (2). We note however, that as in Glauber theory for scattering on a composite target, the elimination of $V_{NN}$ in favour of scattering parameters removes a non-relativistic concept and enables a relativistic generalization.

We still need a prescription for replacing the NR scaling variable in (4c) by a relativistic analog and we chose [16]

$$y^{nr} \to y = (M\nu/q)(1-x-\langle\Delta\rangle/M) \tag{9}$$

with $\langle\Delta\rangle$ an average separation energy per nucleon. In order to reach a suitable generalization of the NR $w^{N/A}$ we invoke sum rules, to be fulfilled by any PN structure function, e.g.

$$\int_0^A dx f^{N/A}(x,Q^2) = \int_{-\infty}^{\infty} dy \phi^{N/A}(q,y) = 1, \tag{10}$$

where we have extended the lower (elastic) limit $y = -q/2 \to -\infty$. Using the proper Jacobian relating the integrands above, we deduce the following correspondence

$$\phi(q,y) \to f^{N/A}(x,Q^2) = \left(\frac{M\nu}{q}\right)\left[\frac{1+4M^2x/Q^2}{1+4M^2x^2/Q^2}\right]\phi^{N/A}\left(q(x,Q^2),y(x,Q^2)\right) \tag{11}$$

with $\nu = Q^2/2Mx$, $q = \nu\sqrt{1+(2Mx/Q)^2}$.

Eq. (11) based on (5) and (8) is the cornerstone for our calculations. It requires as input the above-mentioned $NN$ scattering parameters for the $q$-ranges probed in the 89-008 experiment (Table I). We recall here the appreciable variation of $\sigma_q^{tot}$ in the range of lab momenta $q \approx$ 1-1.5 GeV, getting smooth only beyond $\approx 1.5$ GeV (see for instance Table I in [17]). Except for the 'anomalous' range above, FSI effects decrease slowly and smoothly with increasing $q$ as does $\sigma_q^{pN}$. This is in contrast with the rigorous $1/q$ GRS prediction for regular, static interactions and similar approximate behaviour in regions where effective $NN$ scattering produce diffractive scattering, smooth in $q$ (see the analysis for the strikingly similar case of the structure function or response of liquid $^4$He [18]).

With computed total nuclear structure functions, we approach the inclusive cross section per nucleon (1) which we split in NE and NI contributions where the nucleon retains its identity or becomes excited

$$d^2\sigma_{eA}^{tot} = d^2\sigma_{eA}^{NE} + d^2\sigma_{eA}^{NI} \tag{12}$$

For the determination of the former one needs the elastic structure functions of (on-shell !) nucleons.

$$W_i^{NE}(\nu, Q^2) = \delta(\nu - Q^2/2M)\bar{W}_i^N(Q^2)$$
$$F_2^{NE}(x, Q^2) = \delta(1-x)\bar{W}_2^N(Q^2) \tag{13}$$

Using $\eta = Q^2/4M^2$ and for $\bar{W}_i^N(Q^2)$ the standard combinations of static electric and magnetic form factors $G_i(Q^2)$ one has

$$F_2^{A,NE}(x, Q^2) = xf^{N/A}(x, Q^2)\left\langle\frac{G_E^2(Q^2) + \eta G_M^2(Q^2)}{1+\eta}\right\rangle \tag{14}$$

The corresponding NE of the total inclusive cross section thus reads

$$\left(\frac{d^2\sigma_{eA}}{d\Omega d\epsilon'}\right)^{NE} = \left\langle\frac{d^2\bar{\sigma}_{eN}}{d\Omega d\epsilon'}\right\rangle^{el} w^{N/A}(q, y) = \left\langle\frac{d^2\bar{\sigma}_{eN}}{d\Omega d\epsilon'}\right\rangle^{el}(M/q)\phi^{N/A}(q, y)$$
$$\left(\frac{d^2\bar{\sigma}_{eN}}{d\Omega d\epsilon'}\right)^{el} = \sigma_M\left\langle\frac{G_E^2(Q^2) + \eta G_M^2(Q^2)}{1+\eta} + 2\eta G_M^2(Q^2)\tan^2(\theta/2)\right\rangle \tag{15}$$

and appears as a product of the nucleon-averaged inclusive cross section and the PN structure function $w^{N/A}$ ($f^{N/A}$), Eq.(2). We have emphasized before that the single nucleon momentum distribution $n(p)$ there (cf. (5)), as well as the relative neutron excess resulting in proper $(p, n)$ weighting in (8) and (15), lead to weak $A$ and $Q^2$-dependence of $f^{N/A}(x, Q^2)$ and consequently weak $A$-dependence of $F_2^A(x, Q^2)$ per nucleon [13]. This has two immediate consequences for the nucleon-elastic parts. The first is target specificity and implies ($A_i \geq 12$) (see Fig. 3a in [13])

$$\left[\frac{1}{A_1}\frac{d^2\sigma^{NE}_{eA_1}(\epsilon,\theta)}{d\nu d\Omega}\right] \Big/ \left[\frac{1}{A_2}\frac{d^2\sigma^{NE}_{eA_2}(\epsilon,\theta)}{d\nu d\Omega}\right] \approx 1 \qquad (16)$$

The second one addresses $q(Q^2)$ of $\phi^{N/A}(q,y)$ ($f^{A/N}(x,Q^2)$) which in addition is approximately independent of $A$ and $q(Q^2)$ (see Fig. 2a in [13]). It amounts to approximate universal scaling of $\phi, f^{N/A}$.

Not all of the above holds for the NI part. In a previous calculation we have followed Benhar et al [19,20], who computed that component in the PWIA, disregarding FSI. However, the convolution model embodied in (3) features the same PN function $f^{N/A}$ for both NE and NI contributions. Since a convolution is involved instead of a multiplication, the approximate $Q^2$ independence of $f^{N/A}$ does not immediately reflect on the $Q^2$ dependence of $F_2^A(x,Q^2)/F_2^N(x,Q^2)$. As a consequence, of the two observations above, only (16) remains correct, thus for the NI and total cross sections

$$\left[\frac{1}{A_1}\frac{d^2\sigma^{N,tot}_{eA_1}(\epsilon,\theta)}{d\nu d\Omega}\right] \Big/ \left[\frac{1}{A_2}\frac{d^2\sigma^{N,tot}_{eA_2}(\epsilon,\theta)}{d\nu d\Omega}\right] \approx 1 \qquad (17)$$

We checked the ratios (16) and (17) over the entire $\nu$-range of the SLAC NA3 data [2], i.e. in the region of the quasi-elastic peak, as well as on the high-$\nu$ side of it, which is dominated by $d^2\sigma^{inel}$. We found those that those ratios for $A_1 \geq 12$ are 1 within less than 20% and frequently much less. Of course, the above evidence thus far supports, but does not prove the ansatz (3). In any case, it will be intriguiging to see the outcome of this test for the TJNAF 89-008 experiment.

We conclude this section by mentioning approximations for $F^A$, all based on the high-$Q^2$ limit of the PN structure function (4b) and (11), without touching the $Q^2$ dependence of the nucleon structure function $F^N$ in the convolution (3)

$$f^{N/A}_k(x,Q^2) = (4\pi^2)^{-1} M \int_{|y_k|}^{\infty} dp\, p\, n(p) \qquad (18)$$

The various approximations are distinguished by the scaling parameter $y_k$

$$y^W = \frac{M}{\sqrt{1+4M^2x^2/Q^2}}(1-x-\langle\Delta\rangle/M) \qquad (19a)$$

$$y^{as} = M(1 - x - \langle\Delta\rangle/M) \tag{19b}$$

$$y^{PWIA} = \bar{y}_0\left(1 + \frac{\delta y_0}{\bar{y}_0}\right)$$

$$\bar{y}_0 = -q + \sqrt{(\nu - \langle\Delta\rangle)^2 + 2M(\nu - \langle\Delta\rangle)}$$

$$\frac{\delta y_0}{\bar{y}_0} \approx -\frac{\bar{y}_0}{2A(\bar{y}_0 + q)}\left(1 + \frac{\nu - \langle\Delta\rangle}{M}\right) \tag{19c}$$

with $q, \nu$ expressed as functions of $x, Q^2$ (cf. (11)).

The first approximation we call the West approximation which results when FSI are neglected, i.e. if in (5a) $R(q, y) \to \delta(y)$. In a NR theory it coincides with the asymptotic limit, but in the relativistic generalization there remains a residual $Q^2$ dependence in (19a): Eq. (19b) is the true asymptotic limit. Eq. (19c) represents the high-$Q^2$ PWIA approximation which we shall refer to, when discussing EMC ratios.

### III. PREDICTIONS FOR TJNAF 89-008.

The TJNAF 89-008 experiment covers a much larger range of momentum transfers than the older SLAC NA3 experiment [2] and as a first step, using Eqs (4a), (5)-(9), we had to extend over a wider range of $q, y$ previous calculations of the reduced response $\phi^{N/A}(q, y)$. Details on procedure and input can be found in [13]. Eq. (11) then yields the PN nuclear structure function $f^{N/A}(x, Q^2)$ needed in (3). In view of the above mentioned weak $A$-dependence of $F_i^A$, we focus on one target, namely Fe: results for other species are available.

Figs. 1 show sample results for $f^{N/Fe}(x, Q^2)$ for $Q^2 = 5, 10$ GeV$^2$, computed in the BCA and using three available momentum distributions $n_i(p)$ for that species [13]. Although the transformation $(q, y) \to (Q^2, x)$, in Eq. (11) is not a linear one, the left-hand panels showing $f^{N/A}(x, Q^2)$ naturally resemble the NR reduced responses $w^{N/A}(q, y)$ (cf. Fig. 1c in Ref. [13]). We estimate that the wings of $f^{N/A}(x, Q^2)$ in the BCA, sensitive to ternary and higher order collisions, are approximately located at

$$x \lesssim 0.75, x \gtrsim 1.25; \text{ for } Q^2 \lesssim 3\text{GeV}^2$$

$$x \lesssim 0.60, x \gtrsim 1.60; \text{ for } Q^2 \gtrsim 3 \text{GeV}^2$$

The right-hand panels in Figs. 1 compare the same PN structure functions in the BCA, computed with $n_2$ with the three high $Q^2$-approximations (19). Notice that the unit integrated strength (10) requires all curves to cross: no approximation can for all $x$ lie, either above or below a second one. As expected, differences shrink with increasing $Q^2$.

Next we discuss the nuclear structure functions $F_2^A$, Eq. (3). The NE part has already been mentioned in (14). For the nucleon-averaged NI components we use the parameterizations of $F_2^{p,d}$ given in [21] and interpolated $F_1^{p,d}$ from Bodek et al [24] in combination with $F_i^N \approx F_i^d/2$. As usual it is assumed that those average the contributions from the region of resonances.

We thus make a prediction for $F_2^C$ which has been measured for high $Q^2$, $52 \leq Q^2(\text{GeV}^2) \leq 150$ [4]. For those, the PN structure function has, for any of the cases (19) already reached its asymptotic limit (18). The $Q^2$-dependence of $F_2^C(x, Q^2)$ is then exclusively determined by the same for $F_2^N(x, Q^2)$.

Fig. 2 shows our predictions for $Q^2 = 85$ GeV$^2$ using $n_1(p)$ and the two parameterizations for $F_2^{p,d}(x, Q^2)$ in the relevant $x, Q^2$ region [21]. Here and in the following we retain an average separation energy per nucleon $\langle \Delta \rangle$. It has its origin in the approximate relation between the relativistic $y$-scaling and Bjorken variable. Its value we set at $\approx 0.05$ GeV.

The same predictions result, and those lie somewhat higher than the data for whatever high-$Q^2$ approximation (19) is used. Earlier estimates are in a wide area above the data [25]. At first sight surprisingly large differences result for $x \geq 0.9$ if instead of $n_1$ the distribution $n_2$ is used. This actually agrees with our findings that the largest *relative* changes in the anyhow small $f^{N/A}(x, Q^2)$ for large $Q^2$ occur for $y \approx 0.3$ GeV or $x \approx 0.7$ and $1.3$ (cf. Fig. 1c in [13]). Far smaller differences occur between $F_2^{Fe}(x, Q^2)$, computed with the three $n_i(p)$ available for Fe.

In Figs. 3a,b we show NE and NI components of the structure function $F_2^{Fe}(x, Q^2)$ for $Q^2 = 5, 10 \,\text{GeV}^2$, computed with $n_2$. The elastic part in (3) (cf. (13)) is proportional to $f^{N/A}(x,Q^2)$ and thus decreases rapidly on both sides of the PN Quasi-Elastic peak at $x=1$ (cf. Figs. 1) The NI part decreases with increasing $x$ or (for approximately constant $Q^2$) for decreasing $\nu$, and eventually dominates the elastic part. The size of FSI effects are seen by comparing the BCA with the West-approximation (19a), which neglects those. FSI suppress the nuclear structure function over the entire $x$-range.

Only for the NA18 experiment [3] are there for a range of $Q^2$ some data around $x=1$. We checked either directly or by extrapolation in $Q^2$ that our predictions agree with those.

Having discussed $F_2^A$ we turn to $F_1^A$. There actually is no need for a computation, because one easily shows that in the convolution representation (2)

$$z^A(x,Q^2) \equiv \frac{2xF_1^A(x,Q^2)}{F_2^A(x,Q^2)} = \frac{2xF_1^N(x,Q^2)}{F_2^N(x,Q^2)} = z^N(x,Q^2)$$
$$\lim_{Q^2 \to \infty} z^{N,A}(x,Q^2) = 1 \qquad (20)$$

the predicted Callen-Gross relation ratios $z^A$ for arbitrary $A$ are identical to the *observed* $z^N$ for the nucleon at any $Q^2$ [24]. The above had been noticed by Jaffe in the PWIA approximation for $f^{N/A}$ [8], but as (25) shows, holds for $F_i^A$ in a convolution form (3) with arbitrary $f^{N/A}$.

Before turning to inclusive cross sections we show for orientation in Table I the $x, Q^2$ ranges for each angle and planned energy losses $\nu, q$ [1]. We then display in Fig. 4 inclusive cross sections (1a) on Fe for beam energy of 4 GeV and scattering angles $\theta = 15, 23, 30, 40, 50, 65, 80, 135°$ as function of the energy loss $\nu$ with each cross section sampling an entirely different $x$-range. The calculations correspond to the choice $n_2(p)$ for the single-nucleon momentum distribution and $\langle\Delta\rangle = 0.05$ GeV. The data shown for $\theta = 30°$ are the only ones in NA3 for 4 GeV beam energy [2].

Also shown in Fig. 4 are the NE and NI components of the total nuclear cross sections. The difference in their functional dependence on $x$ (approximately on $\nu$) accounts for the increasing importance and rapid dominance of $d^2\sigma_{e,Fe}^{inel}$ over $d^2\sigma_{e,Fe}^{el}$ for increasing $\theta$. Part of the strong variation with $\theta$ is due to the same in the Mott cross section. Fig. 5 compares predictions for $\theta = 23, 30, 40°$, computed for three distributions $n_i(p)$.

Until the completion of the analysis of the TJNAF 89-009 experiment, we can only compare predictions with the NA3 data for 30° [2]. We first note that the predictions for $\nu \gtrsim 1$ GeV do not distinguish between the different $n(p)$, eliminating there at least one uncertain input element. The agreement is actually very good and even for the deepest inelastic region for that $\theta$, predictions do not deviate more than 40% from the data. With a variation of data over four decades, this cannot be considered a serious disagreement.

One notes a discrepancy also for the lowest energy losses $\nu$ where cross sections through the PN structure function are sensitive to details in $n(p)$ as well as to the energy loss $\nu$. Those do not come as a surprise and have in fact been predicted [13] (see also the discussion after Fig. 1): In the wings of $f^{N/A}(x, Q^2)$ BC and higher order FSI are of the same order of magnitude. At least part of the low-$\nu$ discrepancy may be due to the uncertainty in the knowledge of the single nucleon momentum distribution.

In Fig. 6 we give predictions for Au for $\theta = 23, 30, 40°$ computed with two available momentum distributions for Au. The measure of $A$-dependence is displayed for the ratios of cross sections per nucleon under identical kinematic circumstances. Figs. 7a,b display the Fe/Au ratio for 30 and 40° which show $\leq 20\%$ excursions around 1.0 in agreement with a few NA18 data for the same ratio at $x = 1$ [3]. Those excursions can hardly be discerned in cross sections varying over 3-4 orders of magnitude.

We conclude this Section, commenting on the average separation energy $\langle \Delta \rangle \ll M$, which enters the formalism in the transition (9) from a relativistic GRS-West to the Bjorken variable. The disproportionate, large effects of that quantity $\approx 0.05$GeV on $F^A(x, Q^2)$ for $x \leq 1$ can readily be understood from the convolution (3), where $f^{N/A}(z, Q^2)$ has its maximum for $z = 1 - \langle \Delta \rangle / M$.

In variations of cross sections over several orders of magnitude those are hardly noticed, but the effect becomes prominent when one focuses on relatively small effects. Such is the case for EMC ratios which deviate from 1 by $\lesssim 15\%$. Gurvitz has correctly emphasized that a source of uncertainty may be the very relation (9) which cannot hold for all $x, Q^2$ [26]. The above warns against taking too seriously the accuracy of any prediction for EMC ratios, to which we now turn.

## IV. FSI EFFECTS ON THE NUCLEAR COMPONENT OF TOTAL NUCLEAR STRUCTURE FUNCTIONS IN EMC RATIOS.

EMC ratios scale the total inclusive cross section per nucleon of a charged lepton from a given nucleus against the same for the deuteron. We focus on the 'classical' EMC region $0.25 \lesssim x \lesssim 0.9$ for various $Q^2$ [27,28] It is generally believed that in the above $x$-range the deviations of the EMC ratios from 1 are largely due to nuclear effects. The above range delibeartely excludes $x \leq 0.25$, where in addition to nuclear effects, anti-shadowing [29] and pion cloud effects [30] play a role.

Since the EMC kinematics probes primarily $F_2^A$, the EMC ratio from (1a), (3) reads

$$\mu_A(x, Q^2) = \frac{d^2\sigma^{eA}(x,Q^2)/A}{d^2\sigma^{eD}(x,Q^2)/2} = \frac{F_2^A(x,Q^2)/A}{F_2^D(x,Q^2)/2}$$
$$= \left[ \int_x^A dz f^{N/A}\left(z, Q^2\right) \left\{ \frac{Z}{A} F_2^p\left(\frac{x}{z}, Q^2\right) + \frac{N}{A} F_2^n\left(\frac{x}{z}, Q^2\right) \right\} \right] / F_2^D(x, Q^2) \quad (21)$$

We shall assume that the common ingredient $F_2^A$ in EMC ratios and the TJNAF 89-009 can be treated as for the latter down to $x \approx 0.25$.

We recall the EMC phenomenology [27,28]. With little dependence on, either target mass $A$ or $Q^2$, the EMC ratio for targets between C and Au descends with a slope $\approx$ -0.30-(-0.39) from 1 around $x$=0.28 to a minimum 0.86-0.82, reached at $x \approx$ 0.72-0.76. It then strongly increases, passes 1 around $x \approx 0.82$ and assumes large values when approaching $x$=1 (Fig. 8). The same reasons given above for weak $A$-dependence of $d^2\sigma_{eA}$ hold also for $\mu_A(x, Q^2)$ as does the caution against intuitive reasoning for $Q$-dependence given after (16).

Virtually all studies of EMC ratios have been based on the PWIA for $f^{N/A}$, with the nucleon spectral function as centerpiece. As recalled in Section II one can prove for the PWIA an expression like (2), but with the nucleon structure function $F_2^N$ pertaining to an off-shell nucleon [8–11]. The latter is usually neglected (see however [9]). To our knowledge no additional FSI have been studied beyond the lowest order PWIA, as for instance proposed by Benhar et al for inclusive scattering [19]. Although giving qualitative correct results, the PWIA fails to describe details in the following items:

a) The behaviour of the EMC ratio for $0.18 \lesssim x \lesssim 0.3$ in particular the crossing of $\mu$ around $x \approx 0.28$. Neglect of non-nucleonic and shadowing effects dominating small $x$-values presumably cause that failure.

b) Simultaneous fit for the position and value of the minimum.

c) Detailed behaviour beyond the minimum.

We relegate to the Appendix remarks on the relativistic extension of expansions of (21) and applications of sum rules in order to express coefficients in terms of target observables.

As already emphasized there is no difference between an application to inclusive scattering and the EMC ratio. Using Eqs. (21), (5), (8) and (11) one may establish the size of FSI effects over and above the asymptotic limit of $f^{N/A}$ (and not the PWIA!). As in (12) we decompose

$$\mu_A \equiv \frac{F_2^A}{F_2^d} = \mu_A^{NE} + \mu_A^{N,inel} \tag{22}$$

We start with the dominant NI part. Table II shows global results from the BCA including FSI and from the PWIA for two values of the average separation energy $\langle \Delta \rangle$ and finds as follows:

i) The position of the minimum, its value $\mu_{\langle A \rangle}(x_{min})$ and the slope of the approximately straight section of $\mu_{\langle A \rangle}(x)$ for $x \lesssim x_{min}$ are within $\approx 4\%$ limits independent of the single nucleon momentum distributions $n_i(p)$ and of $A$.

ii) The value of the average separation energy remains the dominant source of variation. We could only get reasonable fits for $\langle \Delta \rangle = 0.05$ GeV. Results for $\langle \Delta \rangle = 0.06$ GeV are decidedly inferior.

iii) There remains a small, noticeable dependence on $Q^2 \lesssim 15 - 20$ GeV$^2$. As alredy mentioned, the relatively weak weak $Q^2$ dependence of $f^{N/A}$ does not immediate reflect on $F_2^A$ due to intermediate folding with $F_2^N$ and divison by $F_2^d$.

In principle one should include the apparently disregarded nucleon-elastic part of $\mu$ in

(22) as generated by the corresponding $F_2^N$ and $F_2^A$, Eq. (14)

$$\mu_A^{NE}(x,Q^2) = F_2^{A(N,el)}(x,Q^2)/F_2^d(x,Q^2)$$
$$= xf^{N/A}(x,Q^2)\left\langle \frac{G_E^2(Q^2)+\eta G_M^2(Q^2)}{1+\eta}\right\rangle \Big/ F_2^d(x,Q^2) \qquad (23)$$

As can be read off (cf. Figs. 2) the contributions decrease rapidly with $Q^2$, but are non-negligible for 'moderate' $Q^2$, starting from $x \gtrsim 0.75$ (see Figs. 2) and increase sharply towards $x = 1$. Typical values are $\mu^{NE}(0.80, 5) = 0.100$; $\mu^{N,el}(0.85, 5) = 0.166$; $\mu^{NE}(0.80, 10) = 0.017$; $\mu^{N,el}(0.85, 10) = 0.032$.

We illustrate our findings in Figs. 8 for $\mu^{Fe}$. The SLAC E87 data for $x \lesssim 0.65$ are for $Q^2 \approx 7$ GeV$^2$ with $Q^2$ increasing with $x \gtrsim 0.65 - 0.7$ [31], while E139 [32] and E140 data [33] are averages for $Q^2 \leq 15$ GeV$^2$. The EMC and BCDMC data [34–36] are for much larger $Q$ where FSI are negligible.

Fig. 8 compares the above data with predictions for $Q^2 = 5, 10$ GeV$^2$. The latter are the BSA results computed with $n_2$ and for $\langle\Delta\rangle$=0.05 GeV, including nucleon-elastic contributions for large $x$. The fits are of reasonable quality and actually encompass the entire classical $x$-range.

Although not in doubt in principle, there are not sufficient overlapping low and high $Q^2$ data for $x \gtrsim 0.75$ which clearly show the presence of nucleon-elastic contributions. The E87 data for $Q^2 \approx 7$ GeV$^2$ are systematically higher than the BCDMS data, but also for $x$ outside the reach of nucleon-elastic contributions. Equal uncertainty holds when instead of the $Q^2$-averaged data for $x \gtrsim 0.65$ in Fig. 8 are replaced by the actual E139 data points (Fig. 1a [32]). A similar search for other targets lead only to C/N data. The trend of extrapolated data in Fig. 3b, Ref. [34] is as expected for the nucleon-elastic contribution.

In Figs. 9a,b we compare the BCA predictions for the data with our standard choice of large-$Q$ approximations (21). There are a only few percent deviations until $x_{min}$ with the exception of the West approximation, which shows a definite larger $\mu$ than the BCA or any high-$Q$ approximation. When trying to understand this special behaviour one should bare in mind that the West approximation corresponds to the case of no FSI, but not necessarily large $Q$. It is interesting that increasing $Q^2$ to $Q^2=15$ GeV$^2$ hardly changes the prediction, except for the West approximation. The latter obviously approaches the asymptotic limit slower than the PWIA, but as Fig. 2 indicates, certainly before $Q^2 = 85$GeV$^2$.

We also note information implicit in a sumrule for $F_2^A$. From the convolution model (3) one finds

$$\epsilon^A(Q^2) \equiv \frac{18}{5} \int_0^A dx F_2^A(x, Q^2) = \frac{18}{5} \int_0^1 dx F_2^N(x, Q^2) = \epsilon^N(Q^2), \qquad (24)$$

with $\epsilon^A$ the fraction of the momentum of a nucleus, carried by quarks in the nucleus. $\epsilon^N$ is the same for a nucleon. The above has been derived by Jaffe for $f^{N/A}$ in the PWIA [8], but actually holds for $F_2^A$ in the from of a convolution (3) with arbitrary $f^{N/A}$: The smearing with a purely nucleonic $f^{N/A}$ does not change the momentum fraction carried by gluons in $F_2^N$. The above sumrule also explains the correlation between slope, the position of the minimum $x_{min}$ and the value of $\mu_A(x_{min})$ when as in Figs. 8 $f_{N/A}$ is studied in various approximations.

We shall make only a few remarks regarding approaches and results of other authors. We can be brief as regards PWIA calculations which all use degree of sophistication for the basic spectral function, without adding FSI (see however [37])[1].

Of different nature is the meson-quark coupling (MQC) model of Guichon, Saito and Thomas [12,38]. As already mentioned in Section II, in that model one distinguishes and treats differently interactions of quarks inside one bag and in different bags. The effect of MQC on *two* bags has clear meson-exchange features of an effective $NN$ interaction which is at the basis of our calculation of $f^{N/A}$. Instead the above authors consider the effect of MQC effect on one nucleon in a mean-field approximation. Not surprisingly, the EMC ratio

---

[1] The lowest order PWIA requires the exact spectral function, just as the lowest order term (4b) in the GRS series requires the exact single nucleon momentum distribution. Occasionally the notion of FSI or correlation effects are used for extensions of the spectral function beyond a mean field approximation. Those do not refer to higher order terms in the PWIA series [22,23].

appears again in a form like the PWIA, with mean-field parameters typical for the MQC model.

## V. CONCLUSIONS

We have given above a unified description of inclusive electron scattering data and EMC ratios based on a nuclear structure function $F_i^A$, expressed as a generalized convolution of structure functions of a nucleon $F_i^N$ and $f^{N/A}$ of a nucleus, composed of point nucleons. In contradistinction to the PWIA, the above non-perturbative approach requires $F_i^N$ for a free, on-shell nucleon. Another feature of interest is the $Q^2$-dependent PN structure function which contains Final State Interactions over and above its asymptotic limit. Those FSI are calculable for large, finite $Q^2$.

One of the interesting consequences of the above is a ratio of inclusive cross sections per nucleon for identical kinematic circumstances, predicted to be only weakly dependent on the species involved. This had already been emphasized for the nucleon-elastic component $F_2^{A(NE)}$ for the older data [2] with as a consequence approximate, universal scaling: $f^{N/A}(x, Q^2)$ is nearly independent of $Q^2$ as well as of $A$. For inclusive cross sections per nucleon as functions of energy loss $\nu$ and scattering angle $\theta$ (thus not for fixed $x, Q^2$), the above implies that ratios of cross sections for two nuclei with $A_i \geq 12$ should be approximately 1. The NA3, NA18 data provide ample evidence for the above within $\approx 20\%$, and frequently substantially better. It will certainly be of interest to test this prediction for the TJBAF experiment, covering much larger kinematic ranges. Agreement will indicate the sufficiency to take data on a very limited set of targets with $A \geq 12$.

Application of the approach explains on the above grounds weak $A$-dependence of EMC ratios. It also enables assessment of FSI effects, which for $x \lesssim x_{min}$ improve the overall agreement with data previously obtained.

We also emphasized the role of nucleon-elastic contributions to EMC ratios for low $Q^2$ towards $x = 1$ with only marginal experimental information from low and high $Q^2$ data in overlapping $x \gtrsim 0.70$ regions. In view of uncertainties in modeling the EMC ratios and knowledge of input parameters, it seems presumptuous to calculate a nuclear structure function with an accuracy better than a few percent. We therefore view our discussion of items of EMC ratios primarily as determining relative effects, rather than the absolute ratios.

Our major emphasis is on the inclusive scattering experiments where small details will be lost in predictions over many decades. For the only set of data available at the relevant kinematics we obtain very good agreement and we look forward to learn from a comparison with the forthcoming extensive TJNAF 89-009 data.

## VI. ACKNOWLEDGEMENTS.

The authors thank Shmuel Gurvitz for vigorous, but instructive discussions on the EMC effect and Brad Fillipone for information on 89-008.

## VII. APPENDIX.

Already in the first theoretical discussions of EMC ratios based on a convolution (3) with a PWIA approximation for the PN nuclear structure function $f^{N/A}(z, Q^2)$, it had been realized that for $x \lesssim 0.5$, $\mu(x, Q^2)$ is largely determined by the expansion of $F_2^N(x/z, Q^2)$ in (21) around the argument $z = 1$ for which the PN nuclear structure function $f^{N/A}(z, Q^2)$ peaks [7]. The result for $F_2^A$ in $\mu$ then becomes

$$F_2^A(x, Q^2)/A = \left\langle F_2^N(x, Q^2) + (1-z)xF_2^{'N}(x, Q^2) + \frac{1}{2}(1-z)^2[2xF_2^{'N}(x, Q^2) + x^2 F_2^{''N}(x, Q^2)] + .... \right\rangle_{f^{N/A}}, \quad (25)$$

where brackets indicate averages of $(1-z)^n$ over the PN structure function. This has been shown before for $f^{N/A}$ in the PWIA, but is from (25) seen to hold for general $f^{N/A}$ in (3), (21).

The averages for the the relativistic GRS case are approached by sum rules, which we first consider for NR dynamics with momentum-independent interactions. The relevant ones read

$$\int dy\, \phi(q,y) = 1; \int dy\, y\, \phi(q,y) dy = 0; \int dy\, y^2\, \phi(q,y) = 2m\langle T\rangle/3$$

with $\langle T\rangle$ the average kinetic energy. In order to generalize those for the relativistic analog $f^{N/A}$, we use again (9)-(11). Neglecting for $x \lesssim 0.5$, terms of order $x^n, n \geq 3$ one finds to leading order in $(m/Q)^2$, $(\langle\Delta\rangle/m)^2$ and $\langle T\rangle/m$

$$\begin{aligned}
\langle(1-x)\rangle &= -\frac{\langle\Delta\rangle}{m} + \frac{2m^2}{Q^2} \\
\langle(1-x)^2\rangle &= \frac{2\langle T\rangle}{3m} + \frac{8m^2}{Q^2}\frac{\langle T\rangle}{3m} - \frac{\langle\Delta\rangle}{m}
\end{aligned} \qquad (26)$$

Those have the standard limits for $Q^2 \to \infty$.

**Figure Captions.**

Fig. 1a. The point-nucleon (PN) nuclear structure function $f^{N/Fe}(x, Q^2)$, $Q^2 = 5\,\text{GeV}^2$ for Fe, computed for 3 momentum distributions $n_1$ (...), $n_2$ (—); $n_3$ (– –). The right-hand panel compares the BCA for $n_2$ (—) with approximations: West (19a) (– –), the asymptotic limit (19b) (- -) and the PWIA (19c) (...).

Figs. 1b. Same as Fig. 1a for $Q^2 = 10\,\text{GeV}^2$.

Fig. 2. Total structure function of C for $Q^2 = 85\,\text{GeV}^2$, computed with $n_1$. Legend as for right hand panel of Fig. 1a.

Fig. 3a. The nuclear structure function $F_2^{Fe}(x, Q^2)$ for $Q^2 = 5\,\text{GeV}^2$ for Fe (drawn lines) computed with $n_2(p)$. Dots and long dashes are for NI and NE parts; long dashes correspond to neglect of FSI (West approximation (19a).

Fig. 3b. Same as Fig. 2a for $Q^2 = 10\,\text{GeV}^2$.

Fig. 4. Total inclusive Fe$(e, e')$X cross sections for 4 GeV beam energy, computed with $n_2$. Drawn, dotted and dashed curves are the total, NE and NI components, Eq. (12) and correspond from left to right to scattering angles $\theta = 15°, 23°, 30°, 40°, 50°, 65°, 80°, 130°$. Older data for $\theta = 30°$ are from [2].

Fig. 5. Same as Fig. 3 for $\theta = 23°, 30°, 40°$ for distributions: $n_1$ (...), $n_2$ (—); $n_3$ (- -).

Fig. 6. Total inclusive Au$(e, e')$X cross sections for $\theta = 23°, 30°, 40°$. Predictions are for $n_1$ (—), $n_3$ (- -).

Figs. 7. Ratio of inclusive cross sections per nucleon, $\theta = 30, 40°$, for Fe and Au. $n^{Au} = n_3, n^{Fe} = n_2$ (-) and $n^{Fe} = n_3$ (...).

Fig. 8. The EMC ratio $\mu_{Fe}(x, Q^2)$ computed in the BCA using $\langle \Delta \rangle = 0.05$ GeV. Drawn and dotted curves are BCA predictions for $Q^2 = 5, 10\,\text{GeV}^2$ including the NE component. Data are from [31] (crosses), [32,33] (circles), [34,35] (squares) and [36] (triangles).

Fig. 9a. Same as Fig. 8. For $Q^2 = 5\,\text{GeV}^2$ BCA results (—) are compared with additional large $Q^2$ approximations from (21). Legend as in right hand panel of Fig. 1a.

Fig. 9b. Same as Fig. 9a for $Q^2 = 10\,\text{GeV}^2$.

Table I

| $\theta$ | $\nu$-range(GeV) | $q$-range (GeV) | $x$-range | $Q^2$-range (GeV$^2$) |
|---|---|---|---|---|
| 15 | 0.85-0.30 | 1.26-1.05 | 0.54-1.79 | 0.86-1.01 |
| 23 | 1.31-0.30 | 1.85-1.56 | 0.69-4.17 | 1.71-2.35 |
| 30 | 2.02-0.79 | 2.49-2.02 | 0.56-2.32 | 2.12-3.44 |
| 40 | 2.42-1.42 | 2.97-2.62 | 0.65-1.81 | 2.96-4.83 |
| 50 | 2.92-1.92 | 3.41-3.10 | 0.56-1.65 | 3.09-5.94 |
| 65 | 3.22-2.52 | 3.74-3.63 | 0.60-1.44 | 3.60-6.84 |
| 80 | 3.44-3.00 | 3.94-3.95 | 0.57-1.17 | 3.70-6.61 |
| 135 | 3.59-3.46 | 4.28-4.37 | 0.80-1.09 | 5.39-7.10 |

$x, q, \nu, Q^2$-ranges covered by the eight angles of the TJNAF 80-009 inclusive scattering experiment of 4 GeV electrons for decreasing energy losses $\nu$ in Fig. 3.

Table II

| | $Q^2$= 5 GeV$^2$ | | $Q^2$=10 GeV$^2$ | | $Q^2$=15 GeV$^2$ | | $\langle$Exp$\rangle_A$ |
|---|---|---|---|---|---|---|---|
| $<\Delta>$ (GeV) | 0.05 | 0.06 | 0.05 | 0.06 | 0.05 | 0.06 | |
| $\langle x_{min}\rangle_{A,n_i}$ | 0.79 | 0.72 | 0.61-0.58 | 0.64-0.61 | 0.58-0.55 | 0.61-0.58 | 0.73 |
| | 0.70 | 0.73 | 0.60 | 0.67 | 0.58 | 0.63 | |
| $\langle \mu_{min}\rangle_{A,n_i}$ | 0.84 | 0.79 | 0.85 | 0.80 | 0.85 | 0.81 | 0.86-0.82 |
| | 0.83 | 0.77 | 0.84 | 0.79 | 0.83 | 0.80 | |
| $\langle'$Slope$'\rangle_{A,n_i}$ | -0.30 | -0.38 | -0.34 | -0.42 | -0.36 | -0.44 | (-0.30)-(-0.39) |
| | -0.31 | -0.39 | -0.31 | -0.44 | -0.35 | -0.42 | |

Position and magnitude of minimum of, and slope of approximately straight decreasing section of the EMC ratio $\mu_{Fe}(x,Q^2)$. Results for different $n_i(p)$ have spread $\lesssim 4\%$. Averages are for C,Al,Fe,Au, unless spread exceeds 4%, in which case results are given for increasing $A$. Each pair of entries corresponds to the GRS (upper) and PWIA (lower) results. Data are averaged over $A$.

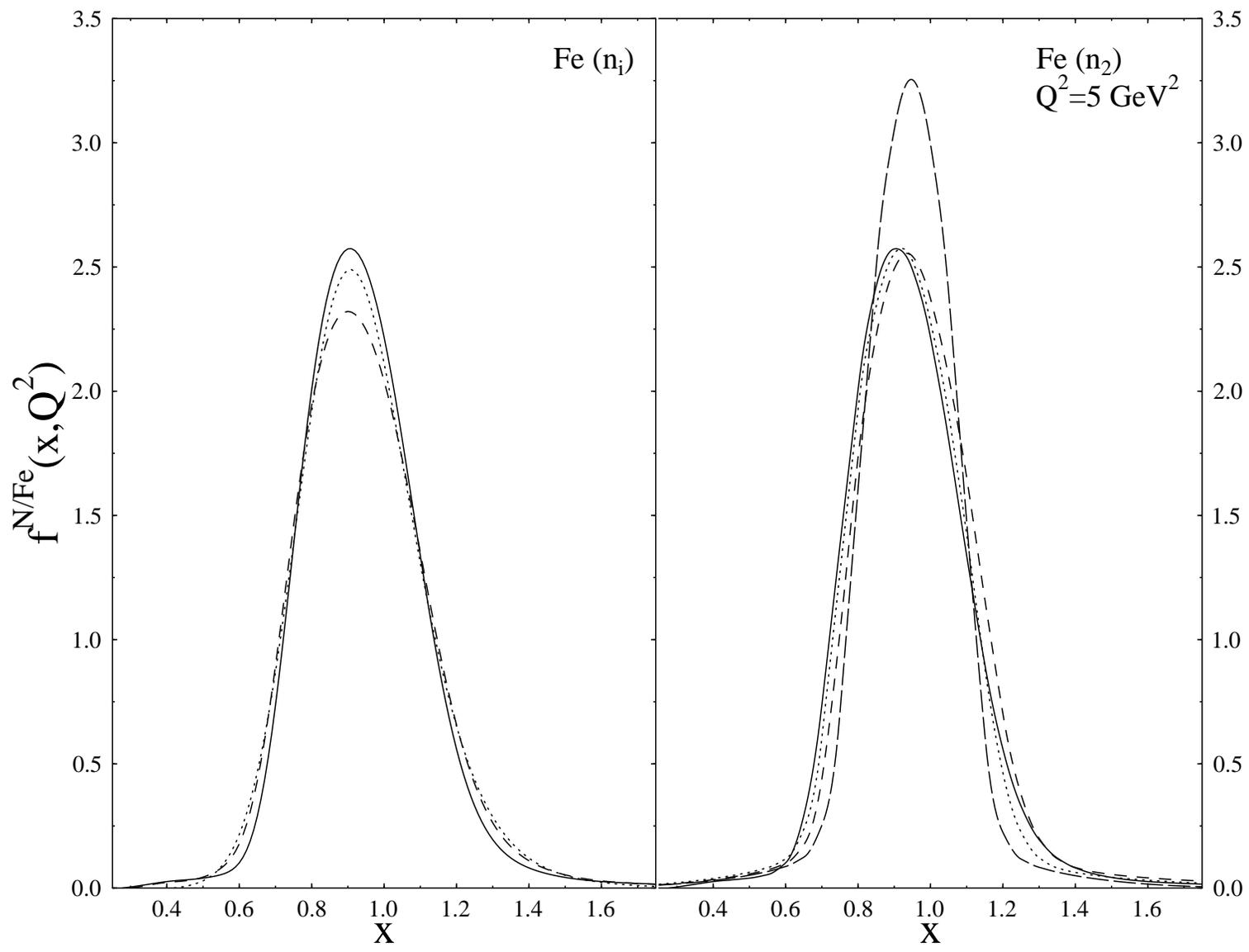

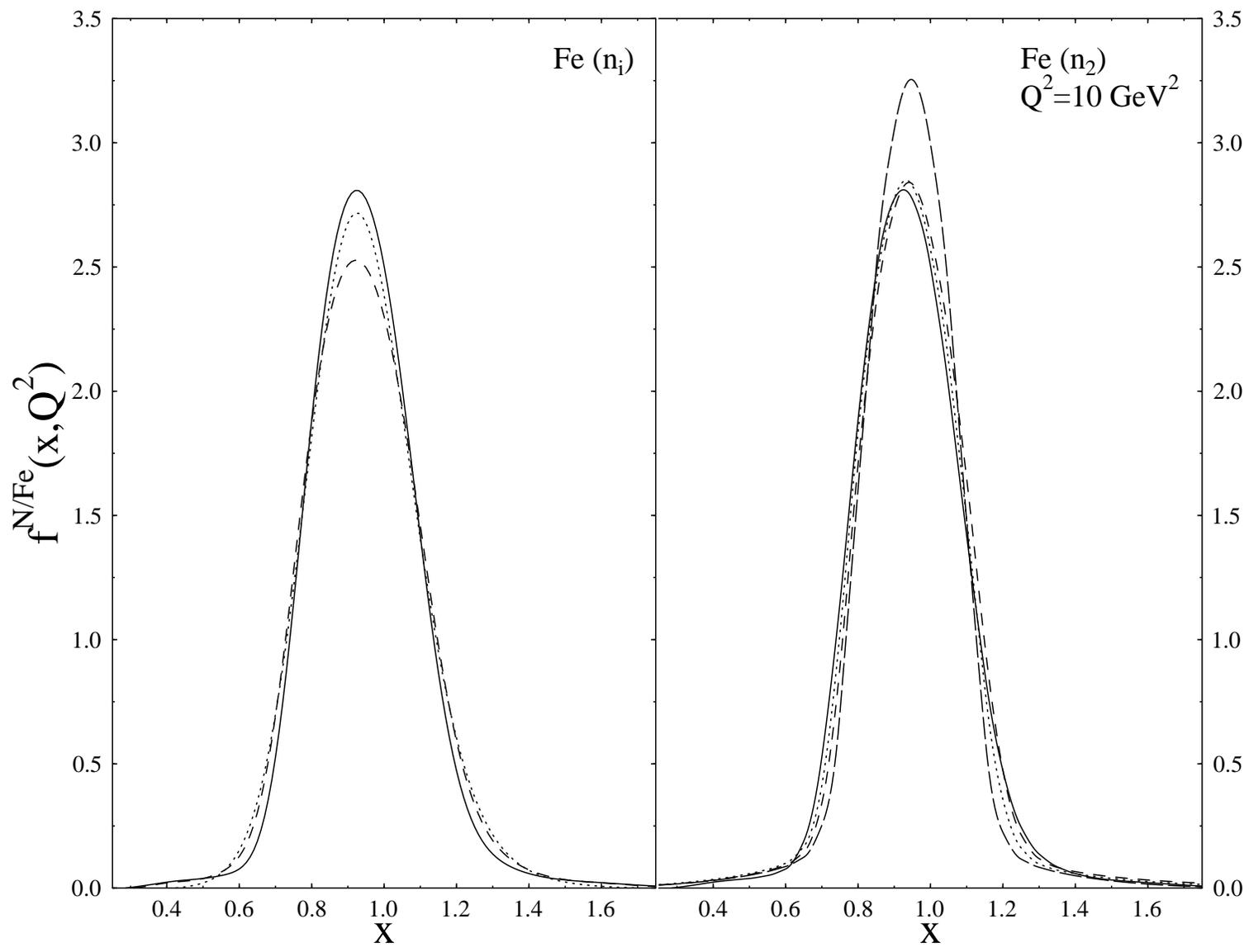

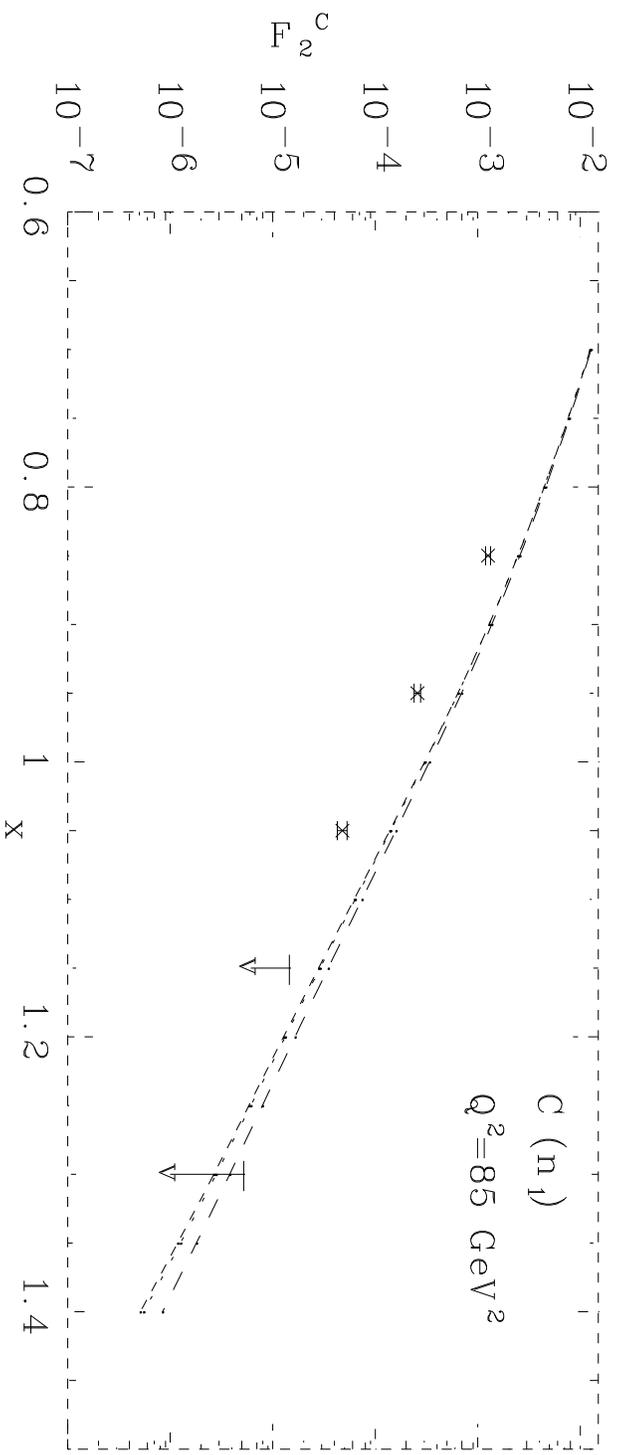

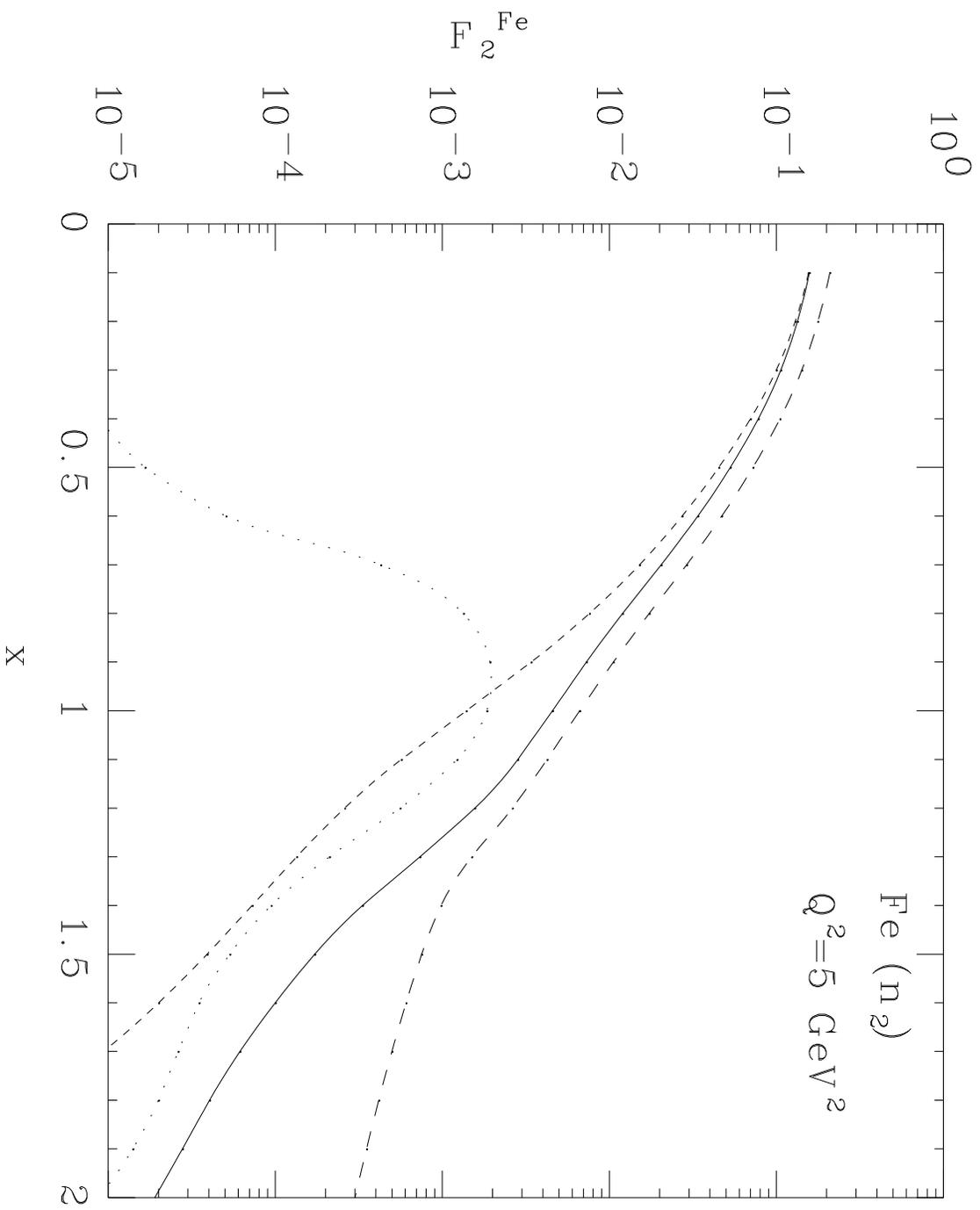

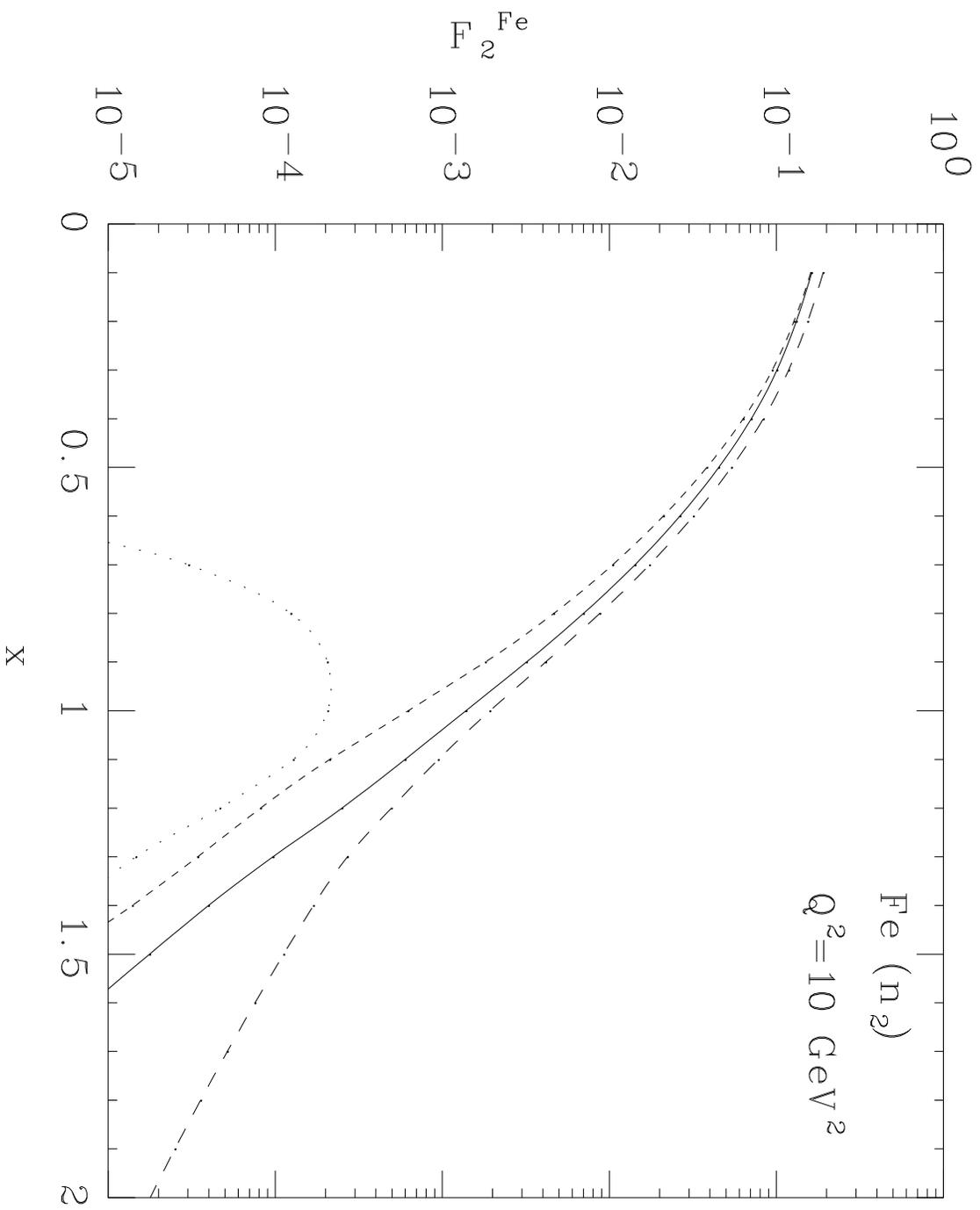

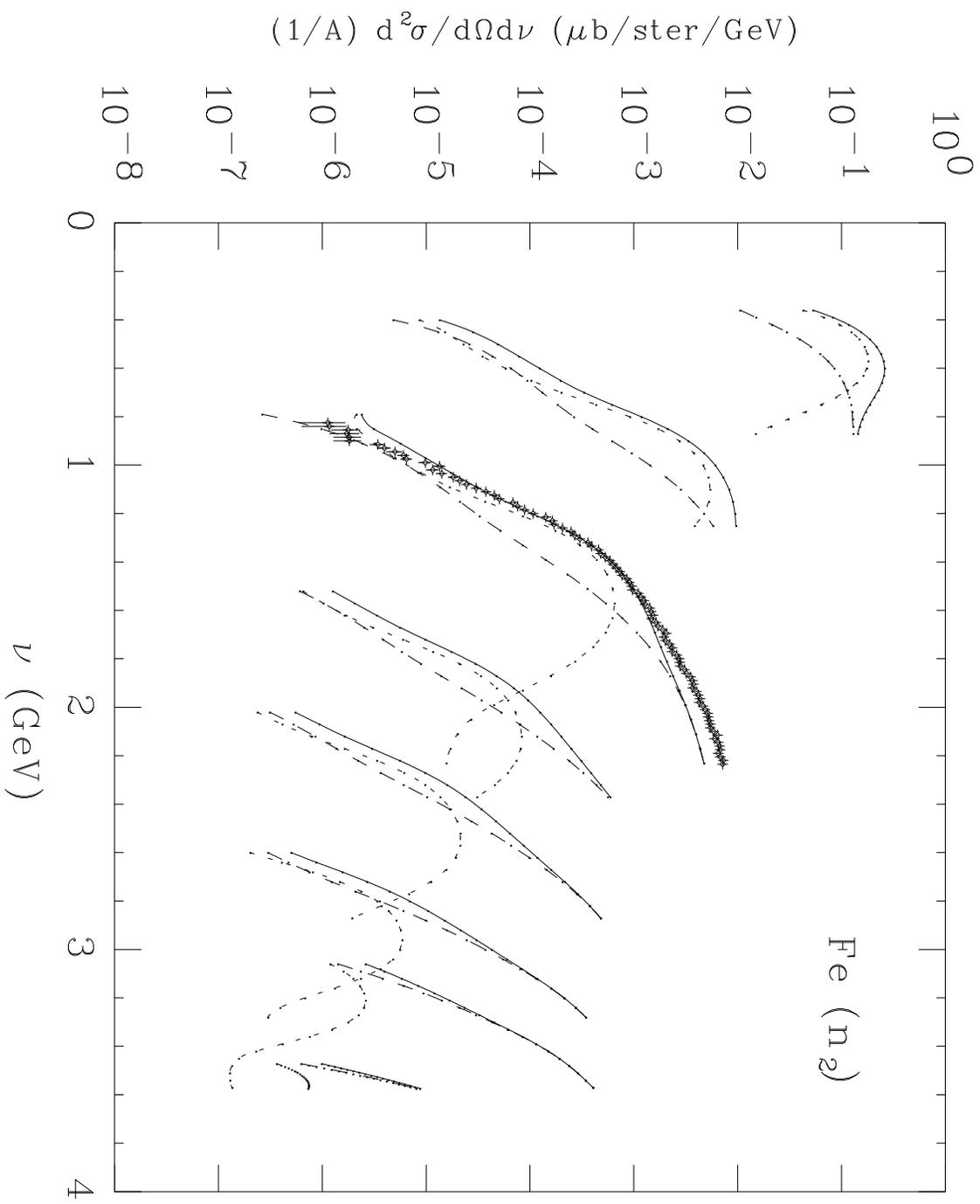

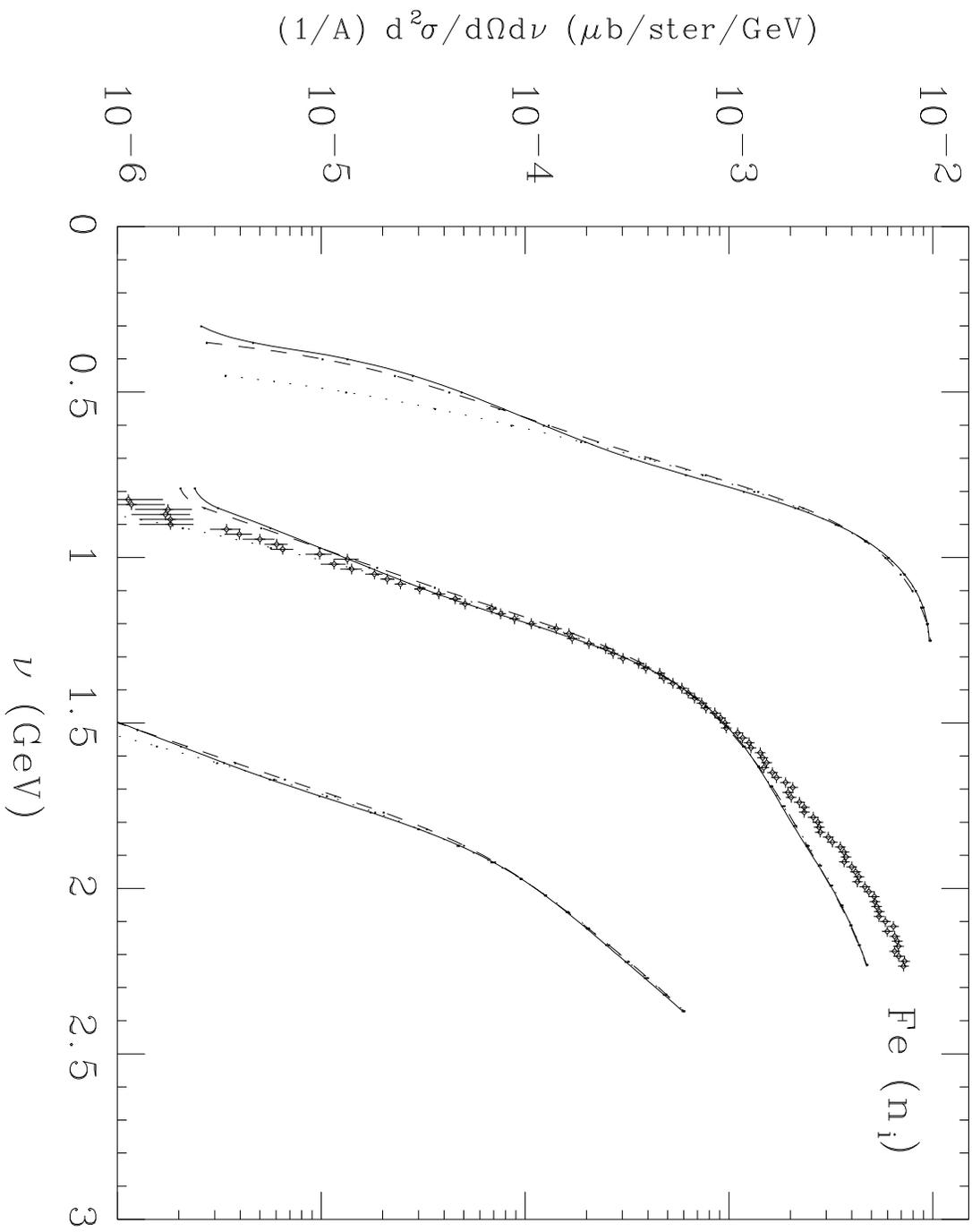

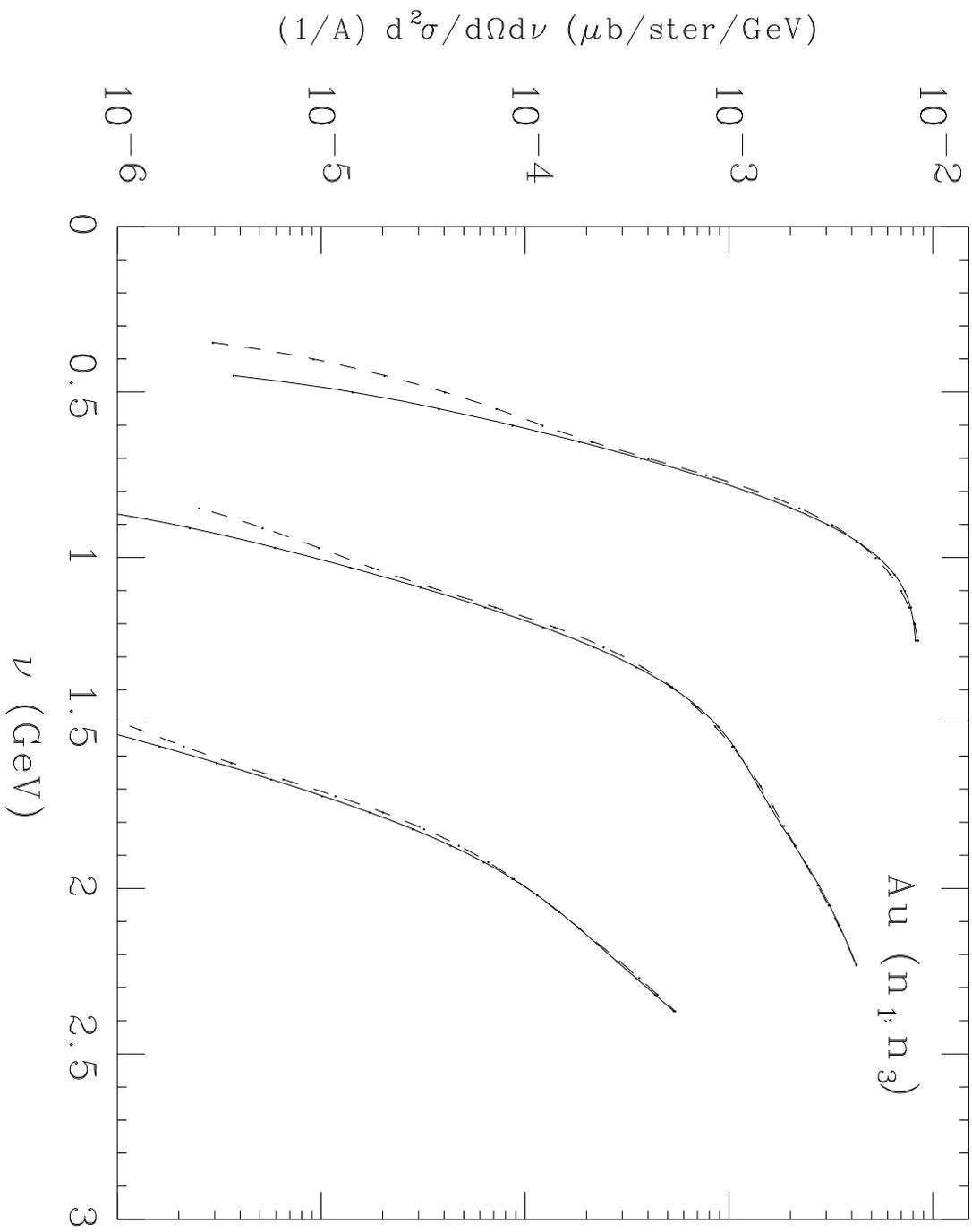

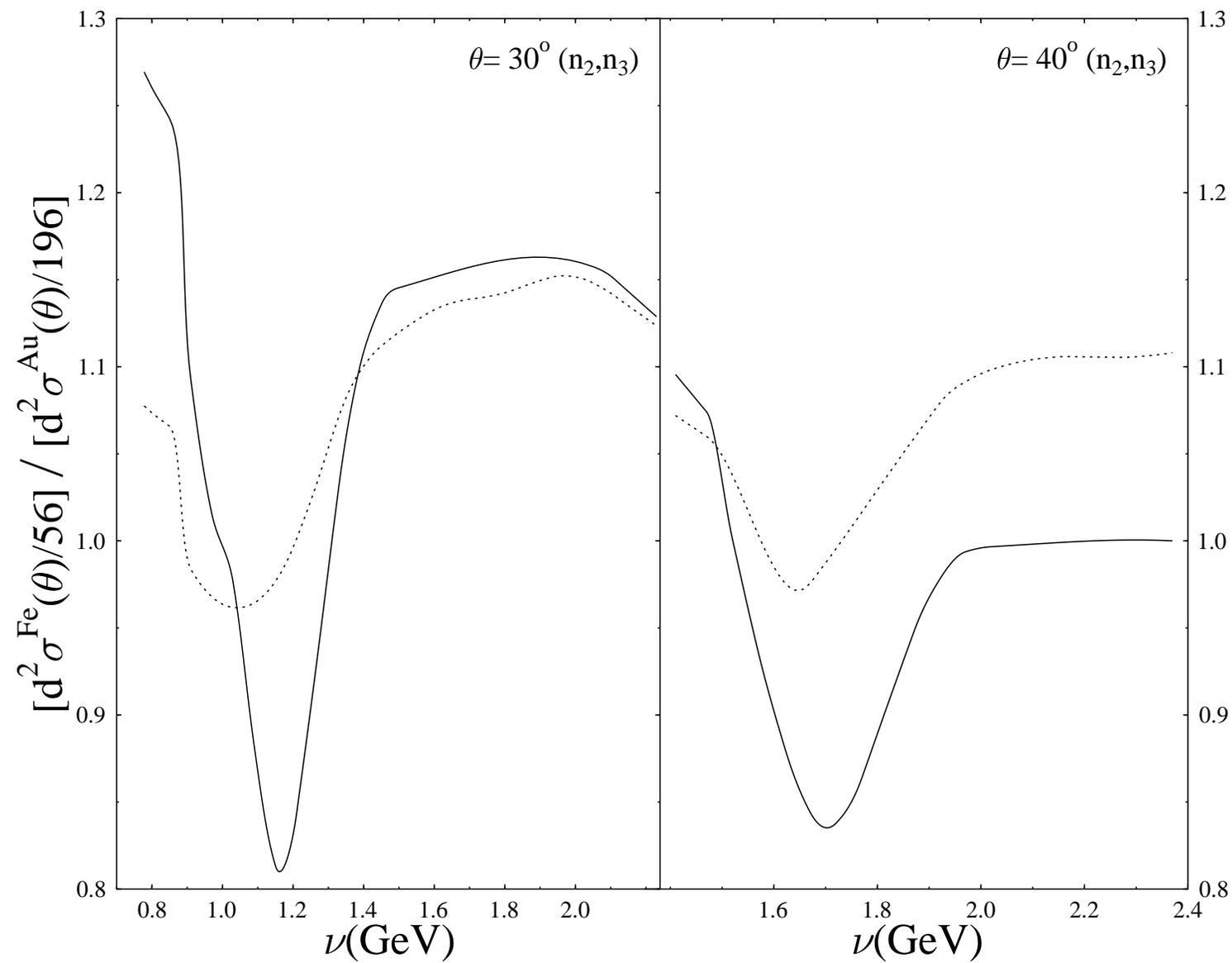

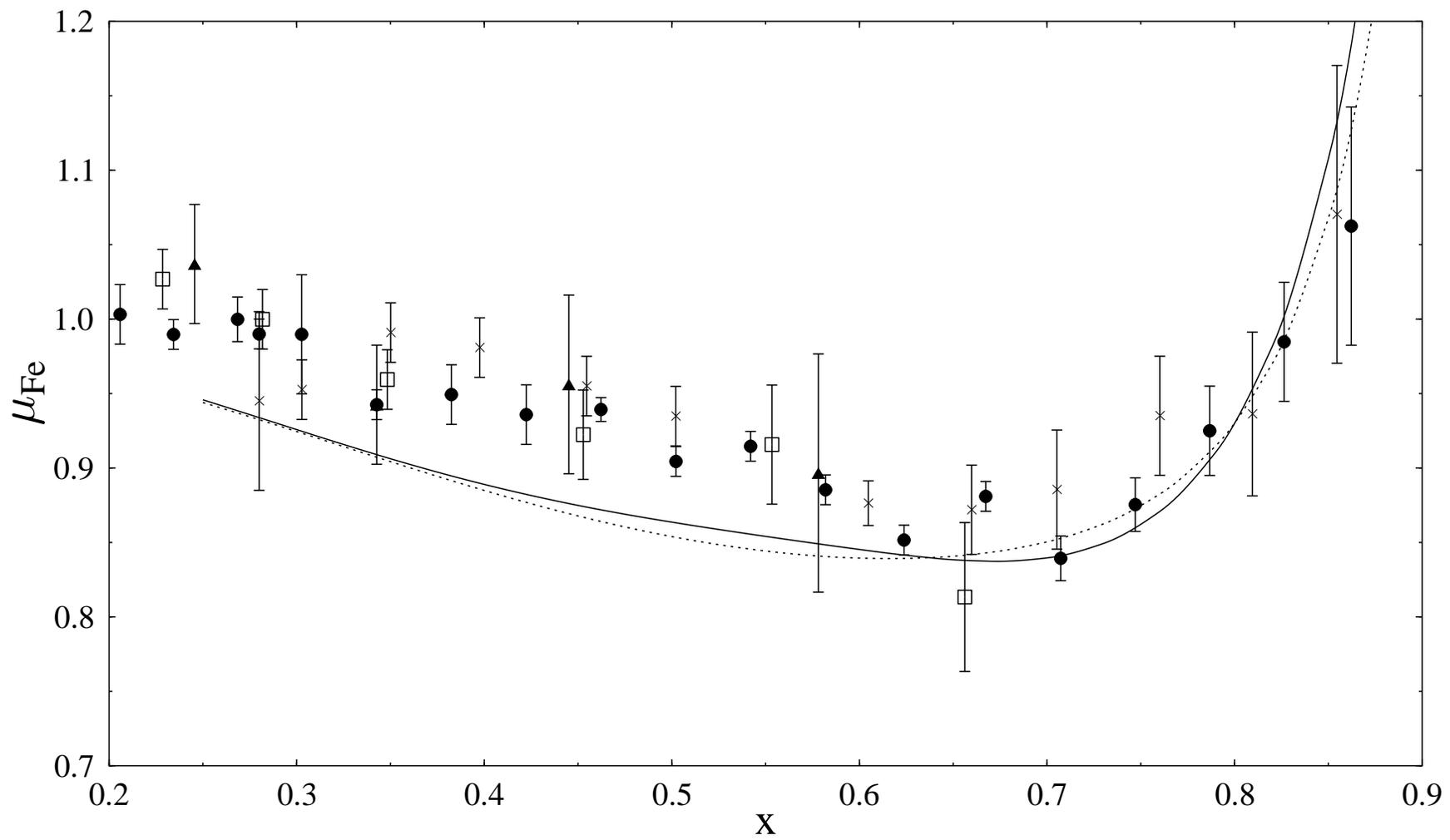

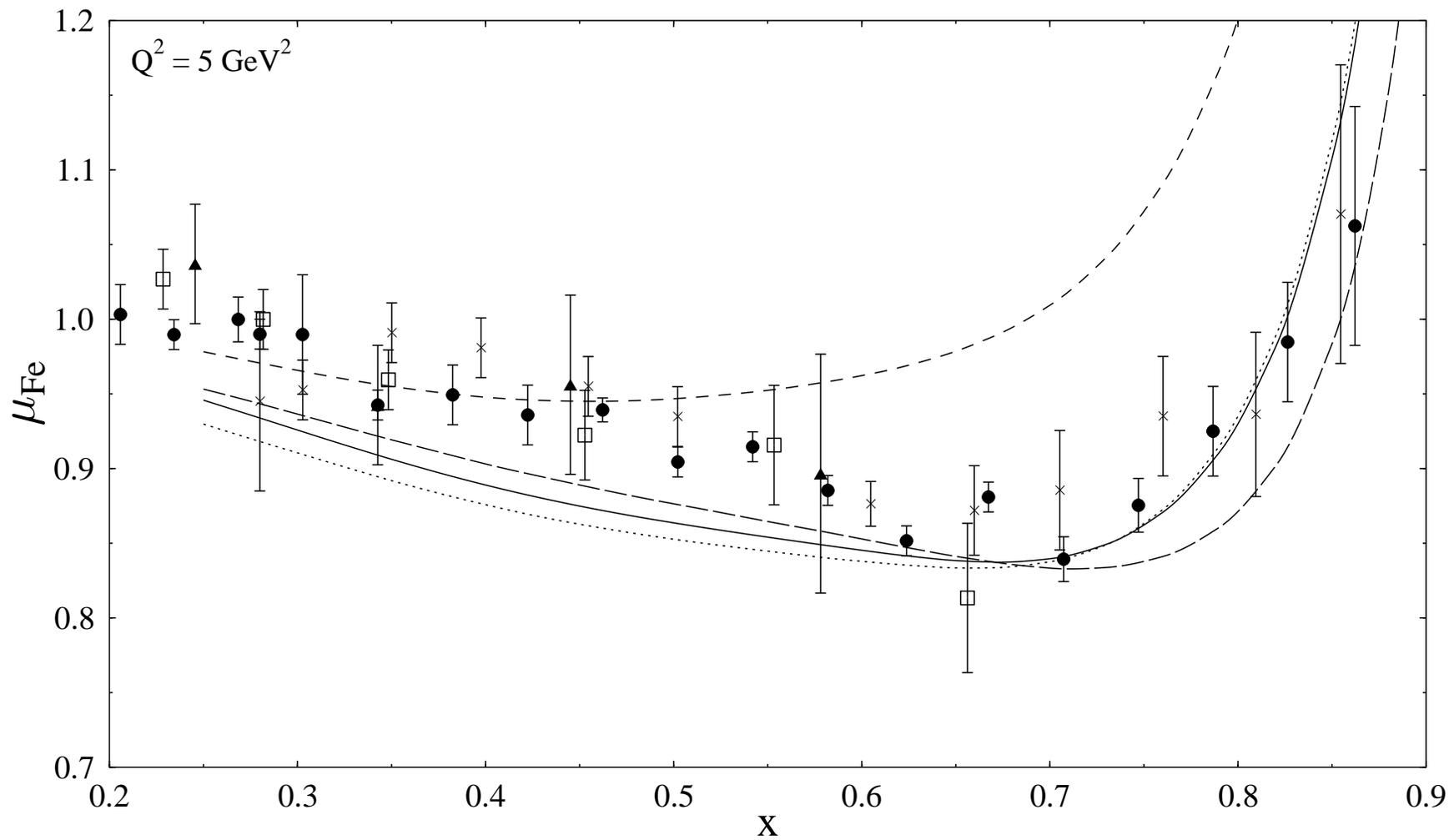

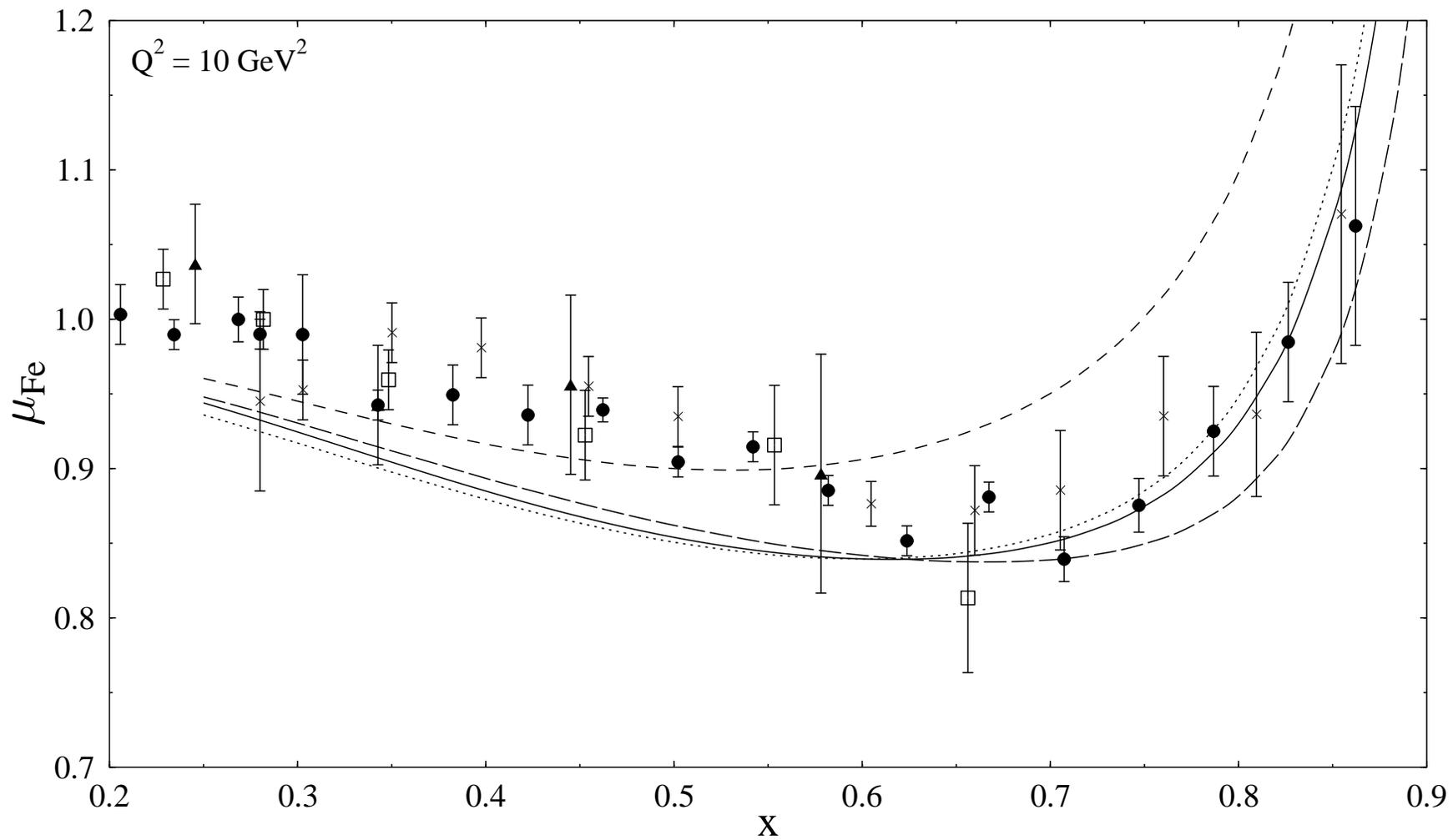